\documentclass[12pt,a4paper]{article}
\pdfoutput=1
\usepackage[utf8]{inputenc}
\usepackage[T2A,T1]{fontenc}
\usepackage[english]{babel}
\usepackage{amssymb,amsfonts,amsmath,mathtext,cite,enumerate,float} %
\usepackage[dvips,pdftex]{graphicx}
\usepackage{geometry} %
\usepackage{ifpdf}
\usepackage{hyperref}
\title{Simulation of an Axial Vircator}

\author{V.~V.~Tikhomirov\thanks{E-mail:vvtikh@mail.ru}, S.E. Siahlo}

\begin{document}
         \maketitle
         \begin{center}
                  Research Institute for Nuclear Problems, Belarusian State University,\\
Bobruiskaya 11, 220030 Minsk, Belarus
         \end{center}

\begin{abstract}
An algorithm of particle-in-cell simulations is described and
tested to aid further the actual design of  simple vircators
working on axially symmetric modes. The methods of correction of
the numerical solution, have been chosen and jointly tested, allow
the stable simulation of the  fast nonlinear multiflow dynamics of
virtual cathode formation and evolution, as well as the fields
generated by the virtual cathode. The selected combination of the
correction methods can be straightforwardly generalized to the
case of axially nonsymmetric modes, while the parameters of these correction methods
can be widely used to improve an agreement between the
simulation predictions and the experimental data.
\end{abstract}

\section{Introduction}

The advancements in the development of high power microwave (HPM) pulsed power sources greatly contribute to the progress in many branches of science and technology. Their most important applications are pulsed heating of plasma, acceleration of
high-current electron or ion beams, feeding of high-power
radar systems, pumping of high-power excimer lasers, destruction
of atmospheric freon to protect the ozone layer. They are
hoped to find applications in the equipment vulnerability testing
as well as in the development of defense systems.

Vircators,  based on the oscillations of a dense electron bunch, called
the virtual cathode (VC), -- are believed to be promising
microwave pulsed power sources with the output power up to 10 GW and
more \cite{1,2}. They have many advantages over more conventional devices, including the simplicity of the hardware design,
compactness, the ability to operate in the absence of an external
magnetic field, the potential for tuning frequency, moderate
requirements to the beam's quality and the simple waveguide
configuration. What is more, they can serve as pulsed sources of
electrons and neutrons. Unfortunately, currently available vircators  have not very high efficiency and unstable generation frequency.

The use of electron beams with currents exceeding the
space-charge limiting current represents the principle virtue of vircators.  Moving downstream of  the anode along of the waveguide, an  electron beam is slowed-down and forms a VC. The growing electric field of the VC reflects the newly incoming electrons, forcing them to oscillate between the virtual and real cathodes. Reflection, mutual repulsion, and the electron drift beyond the VC cause it to destruct. The accompanying VC charge and shape oscillations, together with the electron oscillations are the sources of intense radiation.

That the  electrons motion is strongly nonlinear,
unsteady, and multi-flow allows one to study and design vircators only proceeding from the first principles of electrodynamics and kinetics that are conventionally implemented in the
particle-in-cell (PIC) method that combines the finite-difference
formulation of Maxwell's equations and finite-size particle method \cite{3,4}.
Having no alternative for solving  a wide range of problems dealing
with the application of supercritical currents,  this approach has
been  implemented in a range of computer codes, such as MAGIC
\cite{5} and KARAT \cite{6}.
It is essential that the discrete methods of the solution of Maxwell's equations
cannot be used without regular corrections and smoothing (numerical
filtering ) of numerical solutions for the fields, as well as charge and
current densities.
The methods and techniques for these correction procedures vary greatly and
sometimes lack a serious ground. This limits the possibility to
check and adjust such procedures within the standard codes.
However, holding complete information about the correction and smoothing
procedures, the developer has every possibility to
optimize these procedures and adjust them to specific
applications, as well as to involve experimental data in the code.
This point demonstrates the obvious  need for
self-developed simple codes based on the particle-in-cell method for
the solution of Maxwell's equations.
This paper presents the first version of our self-developed
code enabling  an adequate simulation of the nonstationary,
nonlinear dynamics of supercritical electron beams in the gigawatt-range
axial vircators of the simplest configuration that operate on axially-symmetric modes.

\section{Numerical Solution Algorithm}

\subsection{Finite-difference Scheme and Boundary Conditions}

Let us recall briefly the general approach to the solution of
Maxwell's equations by the particle-in-cell method \cite{3,4,5,6} by
giving specific examples of the applied procedures.
Conventionally, electric and magnetic fields are divided into quasi-static external
fields and rapidly changing "internal" ones that are induced by
the charges and currents of the system. The quasi-stationary
accelerating field is determined by the potential of the
electrodes and can be found by solving  Laplace's equation,
while the internal fields are determined by the boundary
conditions and are obtained from Maxwell's equations.
%\footnote
Following \cite{5,6}, we shall solve Maxwell's equations using the leapfrog method and the Yee mesh comprised of simple rectangular grids (or meshes) shifted in space and time, confining ourselves to the consideration of quite simple constructions
\cite{1,2,6,7,8} with circular waveguides, efficiently
operating on axially symmetric modes $TM_{0i}$ (note that
for coaxial vircators it is crucial that axially assymmetric modes
be taken into account). Our system is described by the following
finite-difference scheme:
\begin{eqnarray}
\label{eq1}
& &\frac{{\left( {E_{r}}  \right)_{i - 1/2,k}^{n + 1}
- \left(
{E_{r}} \right)_{i - 1/2,k}^{n}} }{{c^2\tau} } = \nonumber\\
& & - \frac{{\left( {B_{\theta} } \right)_{i - 1/2,k + 1/2}^{n +
1/2} - \left( {B_{\theta} }  \right)_{i - 1/2,k - 1/2}^{n + 1/2}}
}{{h_{z}} } - \mu _{0} \left( {J_{r}}  \right)_{i -
1/2,k}^{n + 1/2} , \nonumber\\
& & \frac{{\left( {E_{z}}  \right)_{i,k - 1/2}^{n + 1} - \left(
{E_{z}} \right)_{i,k - 1/2}^{n}} }{{c^2\tau} } =
\\
&& \frac{{r_{i + 1/2} \left( {B_{\theta} } \right)_{i + 1/2,k -
1/2}^{n + 1/2} - r_{i - 1/2} \left( {B_{\theta} } \right)_{i -
1/2,k - 1/2}^{n + 1/2}} }{{r_{i} h_{r}} } - \mu _{0} \left(
{J_{z}}  \right)_{i,k - 1/2}^{n + 1/2} ,\nonumber\\
& & \frac{{\left( {B_{\theta} }  \right)_{i - 1/2,k - 1/2}^{n +
1/2} - \left( {B_{\theta} }  \right)_{i - 1/2,k - 1/2}^{n - 1/2}}
}{{\tau} } =\nonumber\\
&& \frac{{\left( {E_{z}}  \right)_{i,k - 1/2}^{n} - \left( {E_{z}}
\right)_{i - 1,k - 1/2}^{n}} }{{h_{r}} } - \frac{{\left( {E_{r}}
\right)_{i - 1/2,k}^{n} - \left( {E_{r}} \right)_{i - 1/2,k -
1}^{n}} }{{h_{z}} }\,\,.\nonumber
\end{eqnarray}

\noindent In these equations $ E_{r} $ and $ E_{z} $ are the radial and longitudinal
 components of the electric field, $ B_{\theta} $ is the azimuthal component  of the magnetic field, and $ J_{r} $ and $ J_{z} $ are the radial and longitudinal  components of the
current density, respectively. The indices $i$ and $k$ in Eqs.(\ref{eq1}) correspond
to the numbers of the grid points (or grid nodes) along the $0r$ and $0z$ axes, whose
coordinates are $ r_{i}=(i-1)h_{r} $ and $ z_{k}=(k-1)h_z $,
where $h_{r}$ and $h_{z}$ are the radial and longitudinal
dimensions of the cells in the $(r-z)$ plane, respectively; $\tau$ is the
time step, and $n$ is the step number.
The quantities $n\pm 1/2, i-1/2$, and $k-1/2$ correspond to the
intermediate time moments $(n\pm 1/2)\tau$ and "half-cell"
coordinates $r = (i-3/2)h_{r}$ and $z = (k-3/2)h_{z}$,
respectively. The time and coordinate steps satisfy the
Courant stability condition
\begin{equation}
\label{CFL}
c\tau < \sqrt {h_{r}^{2} + h_{z}^{2}}\,\,.
\end{equation}

The vircators considered here are simple in design: they consist
of a tubular section and meandres with rectangular cross sections by $rz$ plane \cite{7,8}
(Fig.1);  metal surfaces are parallel to either the $0r$-axis (the emitting surface
 of the cathode, anode mesh, and frontal surfaces of the meandres) or the
 $0z$-axes (outer walls of the diode,  the cathode-holder, the
 resonator and the faces of the meandres). The condition of
 perfect conductivity for these two types of surfaces takes the
 forms $E_{r}=0$ and  $E_{z}=0$, respectively.

The emitted-wave output from the resonator is provided by the
condition
\begin{equation}
\label{eq3}
 c \frac{{\left( {B_{\varphi} }  \right)_{i - 1/2,k + 1/2}^{n + 1/2} + \left(
{B_{\varphi} }  \right)_{i - 1/2,k - 1/2}^{n - 1/2}} }{{2}} =
\left( {E_{r} } \right)_{i - 1/2,k}^{n}
\end{equation}
for the longitudinal index  $k=z_{ex}/h_{z}+1$ corresponding to
the coordinate  $ z_{ex} $ of the output window.
The output power of the vircator is calculated by integrating the Poynting vector $ S=E_{r}H_{\phi} $ over the output window area.

The finite-difference equations (\ref{eq1}) imply that the current
densities $ J_{r} $ and  $ J_{z} $ are specified in discrete
"half-cell" locations of the grid: $(i-3/2)h_{r}$, $(k-1)h_{z}$ and
$(i-1)h_{r}$, $(k-3/2)h_{z}$, respectively.
At the same time, the particle motion defined by the sum of external and internal fields is continuous and can be described by the
Newton-Lorentz equation that includes the  electric and magnetic
fields strengths at continuously distributed points.

  \begin{figure}[h]
       \centering
       % \hspace{-0.0cm} \vspace{0cm}
        \includegraphics[width=0.7\linewidth]{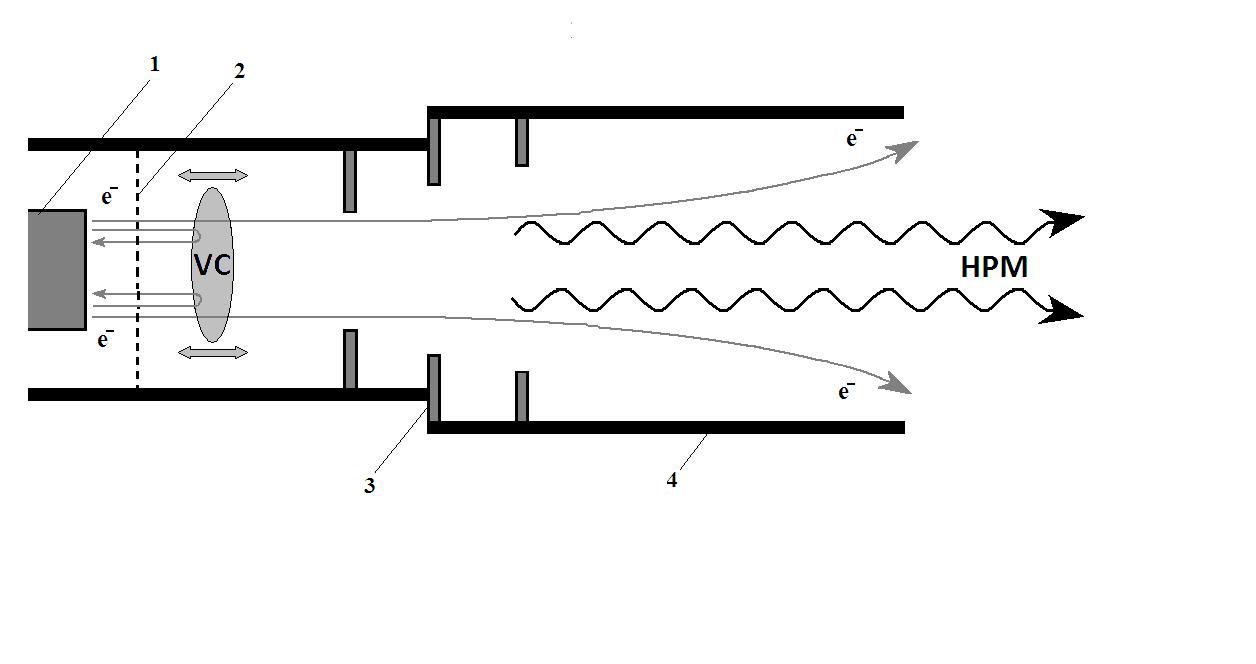}
              \caption{Central longitudinal cross section of the axial vircator: 1 -
cathode, 2 - anode mesh, 3 - meander of the resonator, 4 - side wall
of the resonator. Oscillations of the virtual cathode (VC) together with the oscillations of electrons between the VC and the real cathode, are the sources of high-power microwave (HPM) radiation.}

    \end{figure}

A key to the successful implementation of the PIC method is the
proper direct and inverse conversion  between the physical quantities
defined in a continuous space  and on the discrete grid \cite{3,4}.
In the PIC method, there is  a constant interplay between the grid and the particles as the grid quantities are used to define forces which move the particles, and the particles "send" position and velocity information back to the grid. The conversion between the continuous space and the grid is performed using weight functions that depend on the distance between the particle position and the
coordinate of the grid point.

We have verified the adequate efficiency of the first-order
weighting (also called area weighting) \cite{3,4}, implying that the point charge located at the continuous particle position is allocated to the grid points surrounding the particle by the linear interpolation formulas.
At each time step the weighting procedure is employed twice. First for the interpolation of continuous charge and current densities to
discrete grid points for solving the Poisson (see in what follows) and Maxwell's equations, and second for the interpolation of the
fields, found from Maxwell's equations in the grid points, to
continuously distributed particles.
To achieve a sufficient accuracy of the procedure repeated at each time step, it is essential that the same weight function be used for both direct and inverse interpolations
\cite{3,4}.
It now remains to add that the fields interpolated to continuously
distributed points are applied to find particle
trajectories by solving the Newton-Lorentz equation written in a
time-centered form and solved using the algorithm suggested by
Boris (see \cite{3}).

\subsection{Correction of the numerical solution}

Note should be taken of the fact that the simulation of electron
distribution in a phase space by macro (finite-size) particles,
each containing more than $10^9$ electrons, is a radical step, and one who takes it should be perfectly aware and capable of overcoming consequences.
The problem is that the number of macro particles in a spatial cell is always relatively small, which makes the continuity limit
unachievable even with the maximal computational power. For this
reason, the components of the current density on the right-hand sides of Ampere's law in Eq.(\ref{eq1}) are a permanent powerful source of numerical noise on the scale defined by the cell size of the spatial grid.
Though the processes of interest for simulation  (first and foremost, formation and evolution of a virtual cathode and the related induced fields) manifest themselves over
longer  spatial  scales, the presence of a permanent powerful source of
high-frequency noise inevitably affects the behavior of
macro particles, eventually  causing nonphysical oscillations
at intermediate frequencies.
This particle-induced noise precludes the adequate description of the beam behavior on the most relevant space-time scales and contaminates the physics of interest \cite{3,4,5,6}.

The reduction of this high-frequency noise through filtering the fields enables one not only to obtain comprehensive results with a limited number of macro particles in a cell with long-available computational powers, but even to pass beyond the Courant stability conditions (\ref{CFL}) \cite{5}.
Though justified from the physical viewpoint, filtering is a kind of an artificial trick, and so it is pointless, or at least premature  to discuss the optimal procedures for it.
After years of research, the simulation community actually has developed original approaches to the implementation of filtering, whose  correctness and  efficiency were confirmed by the experimental results \cite{3,4,5,6}.
Based on their experience and governed by the objectives of our team, we use here the filtering procedure that seems simple, reliable, and beneficial in  developing high-power
axial vircators.

Historically, the first problem recognized and solved was the violation
of the Gauss law $\nabla\cdot\mathbf{E} = \rho /\varepsilon _{0} $,
equivalent to the charge conservation equation $\nabla\cdot\mathbf{J} + \partial
\rho /\partial t = 0$ violation.
The most popular and effective method for eliminating this problem is
the Boris correction procedure \cite{3}: finding the
correction potential $\Phi $ that satisfies the Poisson equation
$\Delta \Phi = \nabla\cdot\mathbf{E} - \rho /\varepsilon _{0} $ and replacing
the field  $\mathbf{E}$ by $\mathbf{E} - \nabla \Phi $. Though the approach  is always basically the same, the procedures may vary. For example, the simultaneous over relaxation method for integrating Poisson's equation that is used in the code "KARAT" \cite{6}, is completed for better convergence  by the embedded grid method with Chebyshev acceleration. The adoption of such a highly accurate approach in this context seems redundant, and this is not only because of possible considerable increase of computational time. What really seems redundant is a too high accuracy when finding solutions that give mainly nonphysical information about the current density numerical noise due the use of the PIC method.
These considerations have led us to the application of a much more simple, fast, but adequate procedure: use the  minimal number of iterations of the Jacobi (or point Jacobi) iterative method, the simplest of all iterative methods. One might argue that the Jacobi method has a slow convergence, requiring hundreds and more iterations.
However, tens of thousands of iterations are performed while integrating Eq.
(\ref{eq1}) over time, and so a single pass of Jacobi iteration at each
time step suffices to keep the high-frequency noise,
induced by the PIC method, on the level preventing nonphysical phenomena on the VC characteristic space-time scale.
A similar grounding can well be applied to the Marder's correction algorithm
\cite{9}. Our approach can be considered as a more demonstrative
and simple, but a fully-functional analogue thereof \cite{10}.

Let us note that the simplicity of this approach enables a significant number of iterations of Poisson's equation for the correction potential $\Phi $ at each time step without a noticeable slowing down of the entire simulation procedure.
Usually, the number of iterations can be as large as tens, while
for small spatial areas, such as  the accelerating gap between
the cathode and the anode, this number can even reach hundreds.
In such correction procedure the increasing number of iterations leads to some reduction
in the radiation power and a smother distribution of the fields.
This suggests that the numbers of iterations in different parts of
the vircator construction can also be used as the model parameter to improve the agreement with experimental or some other simulation data.

The correction of the current density instead of a longitudinal
component of the electric field \cite{5} can also compensate for the
deviation from Gauss's law.
At any rate, the correction procedure involves the quantities
specified at the same time moment and is a centered-difference and time-reversible procedure \cite{3}.
Along with electrostatic fluctuations, the errors in the current
density interpolation lead to the excitation of the transverse electric fields,
whose amplitudes grow in time in the absence of attenuation, thus
making increased contributions to the rotors of the electric and
magnetic fields in Eq. (\ref{eq1}).
Earlier for the suppression of short-wave transverse fields, the simplest filtering procedure was used, which means averaging of the filtered quantities over the neighboring grid points with commonly used weights \cite{3,4,6}.
At present, a time-biased (or time-uncentered) semi-implicit scheme due to B. Godfrey \cite{11} is more widely used for filtering particle noise. This procedure implies that $I$ number of iterations ($i=1...I$) of the Ampere and the Faraday equations are performed at each time step and in each point of the grid
\begin{eqnarray}
\label{eq4}
 E^{n + 1,i} & = &  \left( {1 - \tau _{i}}  \right){\kern 1pt} E^{n + 1,i - 1} +
\tau _{i} E^{n,I}  \\
 &  + &  \tau _{i} \delta t\left[ { - J^{n + 1/2}/\varepsilon_{0} +
\nabla \times \left( {aB^{n + 3/2,i - 1} + \left( {1 - a}
\right)B^{n +
1/2,I}} \right)/\left( {\mu_{0} \,\varepsilon_{0}}  \right)} \right],\nonumber \\
 B^{n + 3/2,{\kern 1pt} i} & = & B^{n + 1/2,I} - \delta t\,\nabla \times E^{n +
1,\,i}. \nonumber
 \end{eqnarray}
These iterations result in a selective suppression of the field amplitudes in the upper spectral range; the degree of such filtering, as well as the essential number of iterations, $I$, grows as the parameter $a$ increases from zero to unity. To provide an optimal degree of spectral filtering, the following expression for the iteration coefficients is used \cite{5}:

\begin{equation}
\label{eq5} \tau _{i} = \left\{ 1 + 2\,a\left\{ 1 - cos\left[ {\pi
\left( {i - 1/2} \right)/I} \right]/cos\left( \pi /\left( {2I}
\right) \right)\right\} /\left( {1 - a} \right)^{2} \right\}^{ -
1}.
\end{equation}

Now, the parameters $a$ and $I$ for the correction of
a transverse field are selected by experience. When the stability of
the integration scheme is ensured, these parameters can be further adjusted
to achieve a better agreement with the experiment.

\subsection{Injection of current restricted by space charge  }

As is well known \cite{12,13}, the supercritical currents are generated through explosive electron emission. The magnitude of the current density $J_z$ injected from the cathode is limited by the density of a space charge, whose presence leads to vanishing of the normal component of the
electric field on the emitting surface. The tangential component
$J_r$ there naturally goes to zero too. Under such conditions
Gauss's theorem for the $i$-th grid cell with $0<z<h_{z}/2$ and $0<r<h_{r}/2$ at $i=1$, and $ h_{r} \left( {i-3/2} \right)<r<h_{r} \left( {i-1/2} \right)$ at $ i=2,3,\ldots $ allows one to define the charge values
\begin{eqnarray}
\label{eq6}
 \Delta Q_{i}  =    \pi \varepsilon _{0} h_{r} \left\{
\frac{1}{4}h_{z} \left[ \left( i - \frac{1}{2} \right)\left( E_{r}
\right)_{i-1/2,2} - \left( i - \frac{3}{2}
\right)\left( E_{r}  \right)_{i - 3/2,2} \right]\right.\nonumber\\
\left. +   \left( i - 1 \right)h_{r} \left( E_{z} \right)_{i,1} \right\}-
Q_{i} ,\quad i = 2,3,...\nonumber\\
\Delta Q_{1}  =   \pi \varepsilon _{0} h_{r} \left\{
\frac{{1}}{{8}}h_{z} \left( {E_{r}}  \right)_{1/2,2} +
\frac{{1}}{{4}}h_{r} \left( {E_{z}} \right)_{1,1} \right\} - Q_{1} ,
\end{eqnarray}

\noindent to be injected into these cells to fulfill Gauss's
low at the current time step.
The problem of charges $\Delta Q_{i} $ location inside the
cells does not have an unambiguous solution. Quasi-uniform charge
emission near the cathode surface seems to be quite natural at first
glance.

 \begin{figure}[h]
       \centering
       % \hspace{-0.0cm} \vspace{0cm}
        \includegraphics[width=0.7\linewidth]{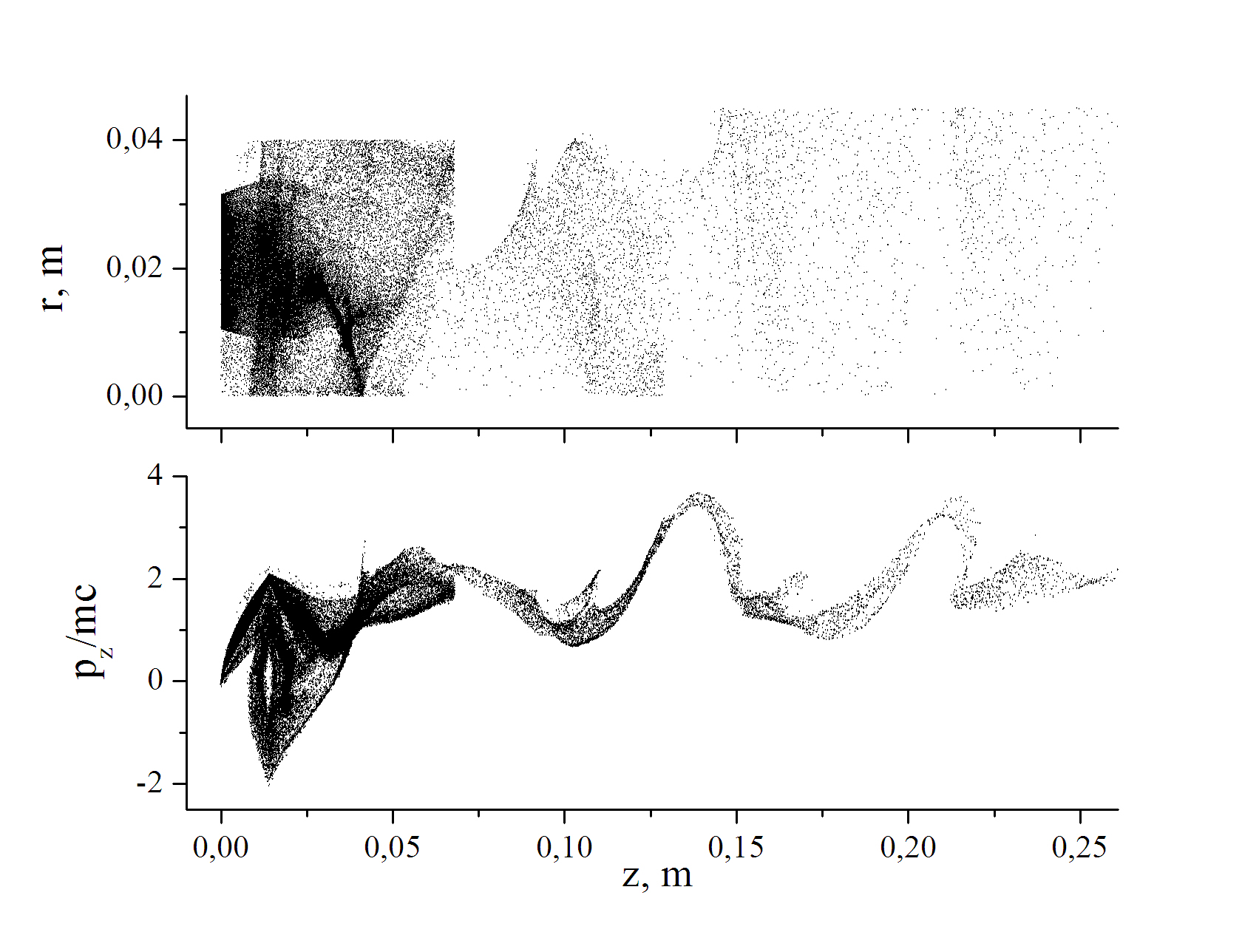}
              \caption{Configurational (top) and longitudinal phase portrait of
the beam in the axial vircator.}

    \end{figure}

However linear extrapolation of the electric field does not provide
adequate values of the accelerating field near the cathode. The small field magnitudes resulting from the vanishing normal component of the field near the cathode surface, lead to
the excessive influence of the field interpolation inaccuracy on the slow particle motion near the cathode surface. Being thus unable to rigorously define the spatial distribution of the
injected electrons, the authors of \cite{14} suggested that the
emitted electrons be placed at the centers $ r_i=h_r (i-1) $, $ i=1,2\ldots$, $ z=h_z/4 $ of
the cells that are contiguous with the cathode surface. Our proposal is also to inject
electrons in the plane $z = h_{z} /4$, however to distribute them quasiuniformly over
$r^{2}$.

\section{Example of Axial Vircator Simulation}

By way of example, we shall consider the axial vircator simulated and tested in \cite{8}. According to Eqs. (\ref{eq1}), under the
conditions of axial symmetry, the electron density oscillations
produce spatially and temporally varying longitudinal and radial
currents. These currents induce the corresponding components of
the electric field both of which in turn induce the azimuthal magnetic
field. Under vanishing tangential components of the electric field
on the walls of a closed cylindrical resonator of radius $R$ and length $L$ $TM_{0il}$ resonator modes \begin{eqnarray}
\label{eq7} E_z & = & E_0J_0(k_{\perp}r)\cos k_z z,\nonumber\\
E_r & = & E_0\frac{k_z}{k_{\perp}}J_1(k_{\perp}r)\sin k_z z,\\
B_{\varphi} & = & iE_0\frac{\omega}{k_{\perp}c^2}J_1(k_{\perp}
r)\cos k_z z\nonumber
\end{eqnarray}
\noindent will be generated, where $k_{ \perp}  = \lambda _{i} /R $, $ \lambda
{\kern 1pt} _{i} $ are the roots of zero-order Bessel
function, $J_{0} \left( {\lambda {\kern 1pt} _{i}} \right) = 0;
\quad k_{z} = \pi l/L, \quad l = 1,2,..., $ $ \omega = c\sqrt {k_{z}^{2}
+ k_{ \bot} ^{2}} $ is the frequency, and  $E_{0} $ is the
amplitude of the wave.

But to provide the radiation output, the  passage is required
between the resonator and the diaphragmatic waveguide that must be
constructed to provide both the effective resonance interaction of
radiation and VC in the resonator and the radiation output from the latter.
It follows from the above that the standing waves (\ref{eq7}) transform into traveling
waves $TM_{0i}$ with the same radial distribution of the intensities.

\begin{figure}[h]
       \centering
       % \hspace{-0.0cm} \vspace{0cm}
        \includegraphics[width=0.7\linewidth]{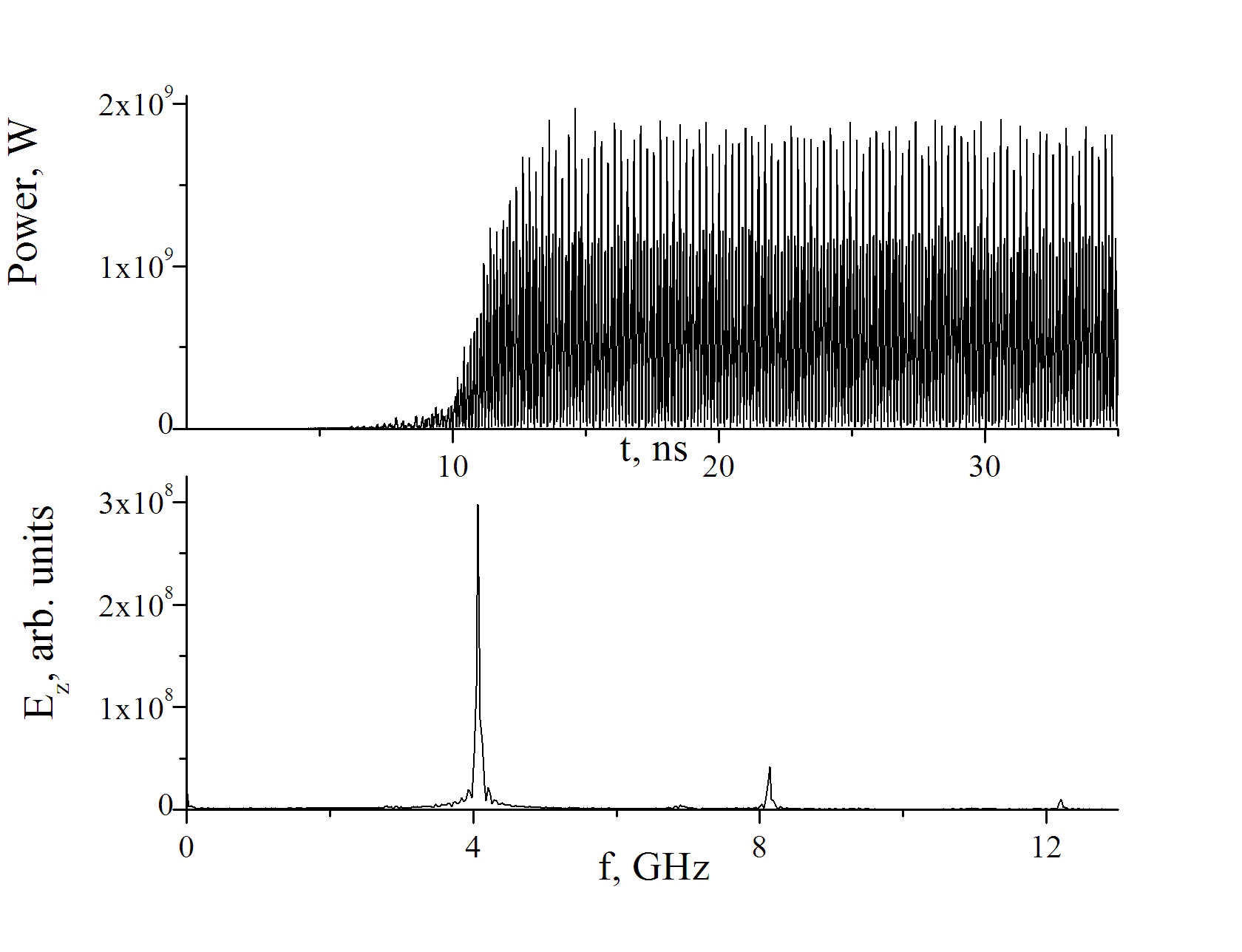}
              \caption{Power vs time (top) and the radiation spectrum (bottom)
 of the axial vircator.}
    \end{figure}

The generation frequency $f=4-4,1$ GHz and the radius $R=4$ cm
of the resonator section of the waveguide enable choosing the
section length $L=5,4$ cm to provide the predominate generation of
the lowest $TM_{011}$ mode.
The dimensions of the other sections of the
waveguide, pictured in Figs. 1 and 3 \cite{8}, were reconstructed
by E.A.Gurnevich; following \cite{8}, the outer and inner radii of
the annular cathode and the cathode-anode gap were taken equal to
32, 12, and 14~mm, respectively, and  the mesh transparency was
taken as $70\%$.
The simulation results for the cathode-anode
voltage of 630 kV are given in Figs. 2 and 3.
Electron distribution in the configurational and the longitudinal
phase space  (Fig.2) illustrate the main features of the VC
dynamics, as well as the significant pulse modulation of the
electrons propagating behind the VC.
As is seen,  the average radiated power is close to 1 GW and the
generation frequency, to 4.1 GHz (see Fig. 3), as reported in \cite{8}.

Let us note that the electrons behind the VC are sometimes
accelerated to the energies exceeding by a factor of two and more
the acceleration energy in the cathode-anode gap, and also have a
rather large energy when they leave the resonator (compare with
\cite{7}).
One can guess that the radiation generated by vircator can be
increased by removing  the electrons from the acceleration process
in the resonator  in the regions, where their energy is minimal,
e.g. using local external fields.

\bigskip

\bigskip

\section{Conclusion}

An algorithm of particle-in-cell simulations was described and
tested to aid further the actual design of simple vircators
working on axially symmetric modes. The methods of correction of
the numerical solution, which were selected and jointly tested,
allow the stable simulation of the  fast nonlinear multiflow
dynamics of the virtual cathode emergence and evolution, as well as
the fields generated by the virtual cathode. The chosen
combination of correction methods can be straightforwardly
generalized to the case of axially nonsymmetric modes, while the
correction methods parameters can be widely used to improve an
agreement between the simulation predictions and the experimental
data.

\section{Acknowledgements} The authors are grateful
to Vladimir Baryshevsky for suggesting the research topic. The
authors also wish to acknowledge the cooperation and useful
discussions with and V.G. Baryshevsky, A.A. Kuraev, S.V.
Anishchenko, A.A. Gurinovich, E.A.Gurnevich, P.V. Molchanov, and
S.N. Sytova.

\end{document}